\documentclass[cits]{PoS}
\usepackage{epsfig}
\usepackage[caption=false]{subfig}
\usepackage{amssymb}
\usepackage{fontenc}
\usepackage{times}
\usepackage{mathptmx}
\usepackage{setspace}
\usepackage{graphicx}
\usepackage{bm}

\title{Fluid dynamics near the QCD critical point}

\ShortTitle{Fluid dynamics near the QCD critical point}

\author{\speaker{Marcus Bleicher} and Christoph Herold\\
        Institut f\"ur Theoretische Physik, Goethe-Universit\"at, Max-von-Laue-Str.~1, 
	60438 Frankfurt am Main, Germany\\
        Frankfurt Institute for Advanced Studies (FIAS), Ruth-Moufang-Str.~1, 60438 Frankfurt am Main, Germany\\	

        E-mail: \email{bleicher@th.physik.uni-frankfurt.de}, \email{herold@th.physik.uni-frankfurt.de}}

\abstract{We present a fully dynamical model to study the chiral and deconfinement transition of QCD simultaneously. 
The quark degrees of freedom constitute a heat bath in local equilibrium for both order parameters, the sigma field and 
a dynamical Polyakov loop. The nonequilibrium evolution of these fields is described by Langevin equations including 
dissipation and noise. In several quench scenarios we are able to observe a delay in the relaxation times near the 
transition temperature for a critical point as well as a first-order phase transition scenario. During the hydrodynamical 
expansion of a hot quark fluid we find a strong enhancement of thermal fluctuations at the first-order transition 
compared to a scenario with a critical point.}

\FullConference{Xth Quark Confinement and the Hadron Spectrum,\\
		October 8-12, 2012\\
		TUM Campus Garching, Munich, Germany}

\begin{document}

\section{Introduction}
The QCD phase diagram and the location of a possible critical point (CP) are subject to intense theoretical and 
experimental research. Lattice QCD calculations support a crossover at vanishing baryon densities \cite{Aoki:2006we}, 
while model studies indicate the existence of a first-order phase transition at high baryon densities ending at a CP 
\cite{Scavenius:2000qd}. In \cite{Stephanov:1999zu} it was shown how to detect the CP in heavy ion collisions by searching 
for divergences in event-by-event fluctuations of transverse momentum or particle multiplicity, an ansatz that has 
recently been refined to higher order cumulants \cite{Stephanov:2008qz, Karsch:2010ck}. However, it is expected that 
finite size and time effects and possibly initial state fluctuations will crucially weaken the expected signals \cite{Bleicher:1998ab,Berdnikov:1999ph}. On the other hand, 
the nonequilibrium evolution during a heavy-ion collision will enhance effects at the first-order phase transition, 
where spinodal instabilities may produce domain formation and clustering in energy and baryon density 
\cite{Sasaki:2007qh,Steinheimer:2012gc, Herold:2013bi}. Hadronization of these clusters will lead to large 
non-statistical fluctuations in the hadron rapidity density within single events, providing an important observable signal 
for upcoming experiments at FAIR and NICA \cite{Mishustin:1998eq}. 

A successful dynamical model to study effects at the QCD phase transition in nonequilibrium has to go beyond usual 
hydrodynamics that includes the phase transition only in the equation of state \cite{Steinheimer:2007iy}. A novel 
approach that includes the dynamics of the order parameters explicitly is given by chiral fluid 
dynamics \cite{Mishustin:1998yc,Paech:2003fe,Nahrgang:2011mg,Nahrgang:2011ll,arXiv:1105.1962}. We recently extended 
this model with the Polyakov loop to consider both the chiral and the deconfinement transition 
\cite{Herold:2013bi, Herold:2013cg}.  

\section{Polyakov-chiral fluid dynamics (P$\chi$FD)}
The basic idea of the model is to explicitly propagate the sigma field and an effective Polyakov loop as the order 
parameters of the chiral and deconfinement phase transition. A fluid dynamically expanding medium of quarks and 
antiquarks provides the locally thermalized background for these fields. This enables us to study relevant effects at 
the CP and first-order transition in a dynamical system of finite size. 

We use the Polyakov loop extended quark meson model \cite{arXiv:0704.3234} with the Lagrangian
\begin{equation}
{\cal L}=\overline{q}\left[i \left(\gamma^\mu \partial_\mu-i g_{\rm s}\gamma^0 A_0\right)-g \sigma \right]q + \frac{1}{2}\left(\partial_\mu\sigma\right)^2 
- U\left(\sigma\right) - {\cal U}(\ell, \bar\ell)~,
 \end{equation}
where $q=(u,d)$ is the constituent quark field, $A_0$ the temporal component of the color gauge field, $\sigma$ the mesonic 
field and $\ell$ the Polyakov loop. The pion degrees of freedom are neglected throughout this work. The potential for the 
sigma field is the usual ``Mexican hat''
\begin{equation}
U\left(\sigma\right)=\frac{\lambda^2}{4}\left(\sigma^2-\nu^2\right)^2-h_q\sigma-U_0~,
\label{eq:Usigma}
\end{equation}
and the temperature dependent Polyakov loop potential is chosen in a polynomial form \cite{arXiv:0704.3234, Ratti:2005jh}:
\begin{equation}
\frac{{\cal U}}{T^4}\left(\ell, \bar\ell\right)= -\frac{b_2(T)}{4}\left(\left|\ell\right|^2+\left|\bar\ell\right|^2\right)-\frac{b_3}{6}\left(\ell^3+\bar\ell^3\right) + \frac{b_4}{16}\left(\left|\ell\right|^2+\left|\bar\ell\right|^2\right)^2~.
\label{eq:Uloop}
\end{equation}
Integrating out the quark degrees of freedom in the partition function ${\cal Z}$ gives us the effective potential:
\begin{equation}
 V_{\rm eff}=-\frac{T}{V}\ln {\cal Z}=U+{\cal U}+\Omega_{q\bar q}~.
\end{equation}
Here, the quark contribution $\Omega_{q\bar q}$ determines the local equilibrium pressure of the quark fluid. In mean-field 
approximation and at zero chemical potential it reads \cite{arXiv:0704.3234}:
\begin{equation}
 \Omega_{q\bar q}=-4 N_f T\int\frac{\mathrm d^3 p}{(2\pi)^3} \ln\left[1+3\ell\mathrm e^{-\beta E}+3\ell\mathrm e^{-2\beta E}+\mathrm e^{-3\beta E}  \right]~.
\end{equation}

\begin{figure}[h]
\centering
   \subfloat[\label{fig:fopot}]{
   \centering
   \includegraphics[scale=0.4, angle=270]{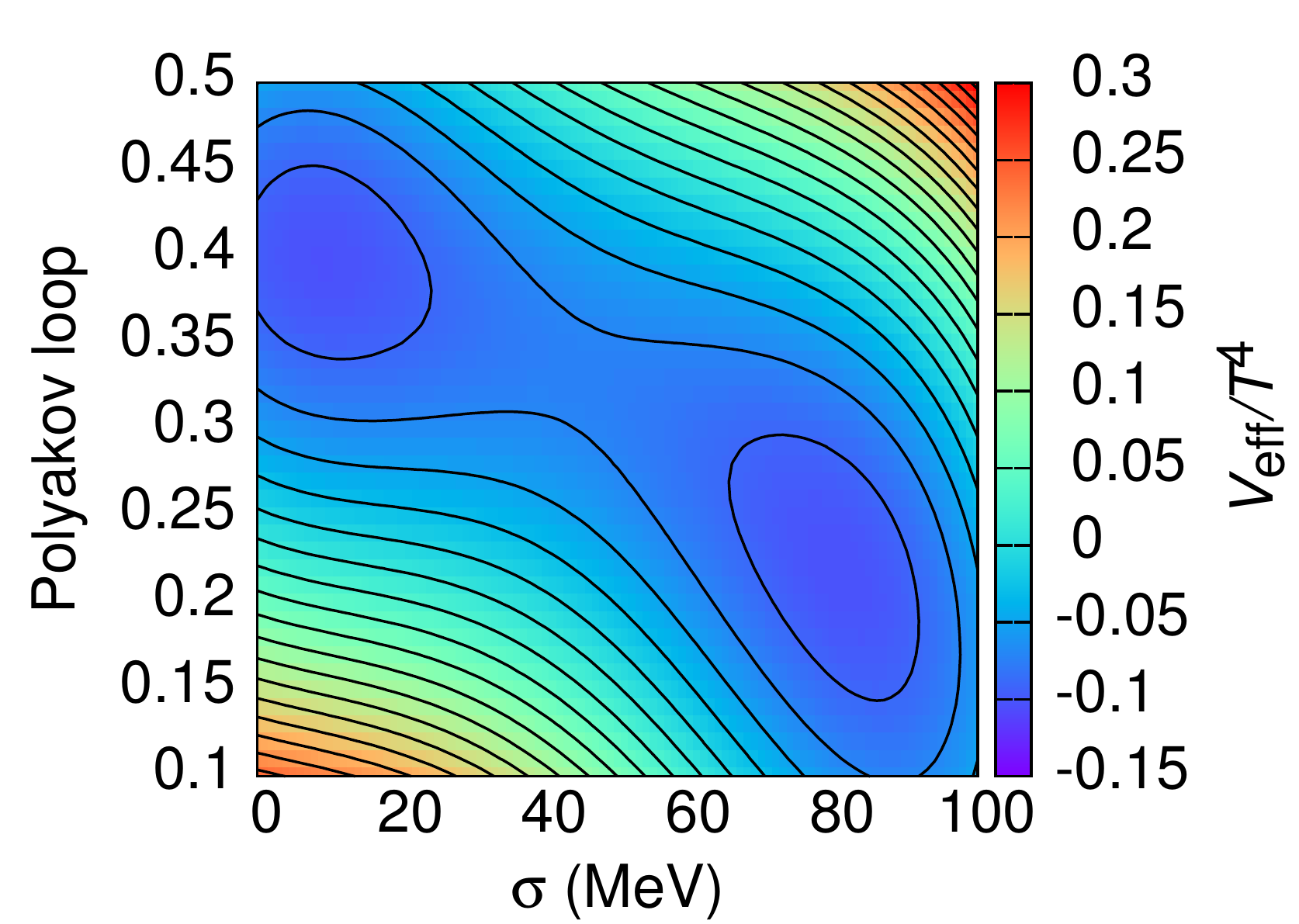}
   }
 \qquad
   \subfloat[\label{fig:cppot}]{
   \centering
   \includegraphics[scale=0.4, angle=270]{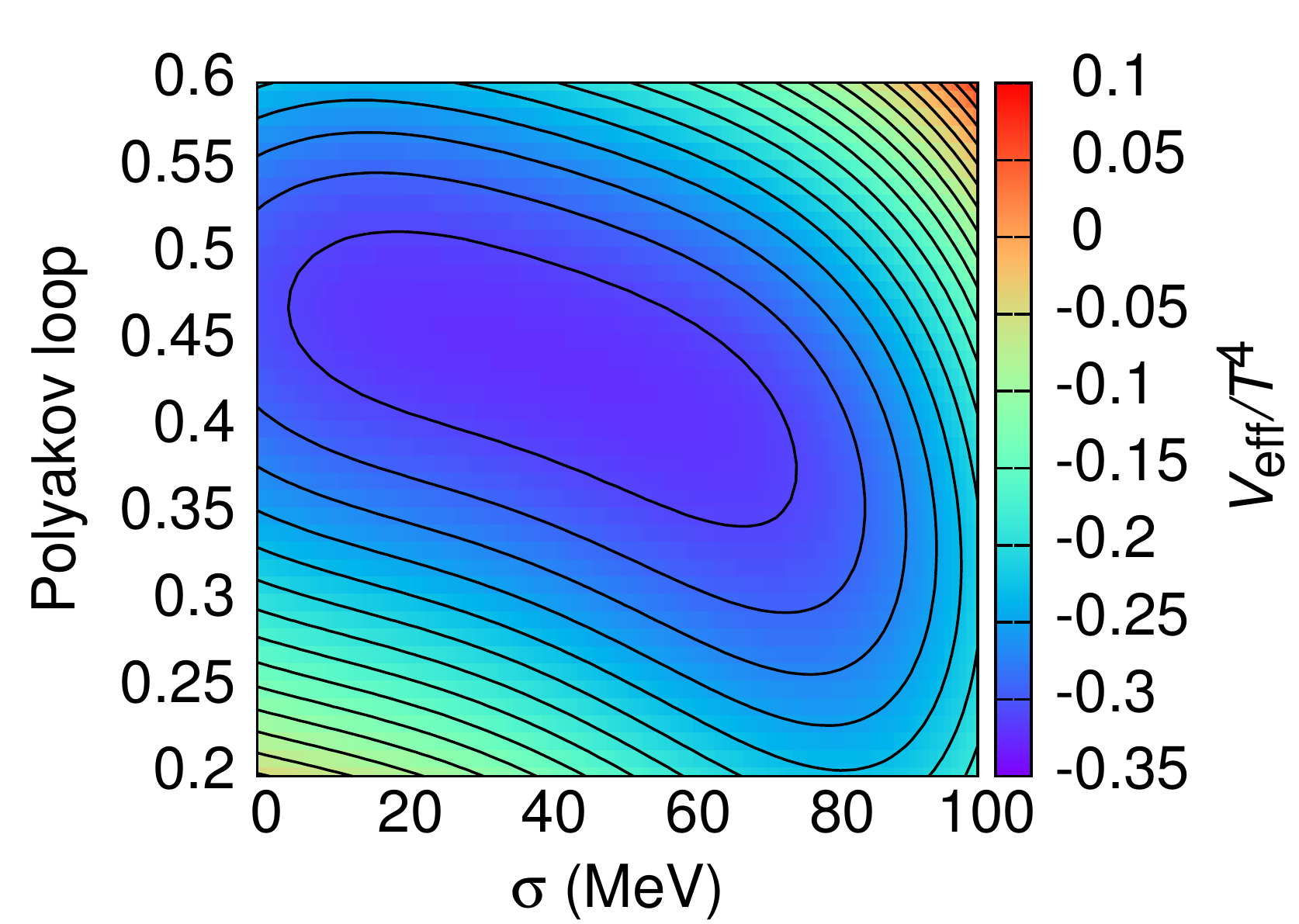}
   }
\caption[.]{\ref{fig:fopot} Effective potential for $g=4.7$, corresponding to a first-order phase transition at $T_c=172.9$~MeV.
\ref{fig:cppot} Effective potential for $g=3.52$, corresponding to a CP scenario at $T_c=180.5$~MeV. 
Both figures are adopted from \cite{Herold:2013bi}.}
\label{fig:potential}
\end{figure}

We tune the strength of the transition by varying the quark-meson coupling $g$. This allows us to study first-order 
phase transitions and transitions through the CP at vanishing baryochemical potential. Figure \ref{fig:potential} shows 
the effective potential for $g=4.7$ (first-order) and $g=3.52$ (CP) at the respective transition temperature
(cf. Ref. \cite{Herold:2013bi}). Note that 
in general one has to choose $g$ such that the product $g\sigma$ resembles the constituent quark mass in vacuum, 
leading to a value of $g\sim 3.3$. 

\begin{figure}[h]
\centering
   \subfloat[\label{fig:susc_sigma}]{
   \centering
   \includegraphics[scale=0.5]{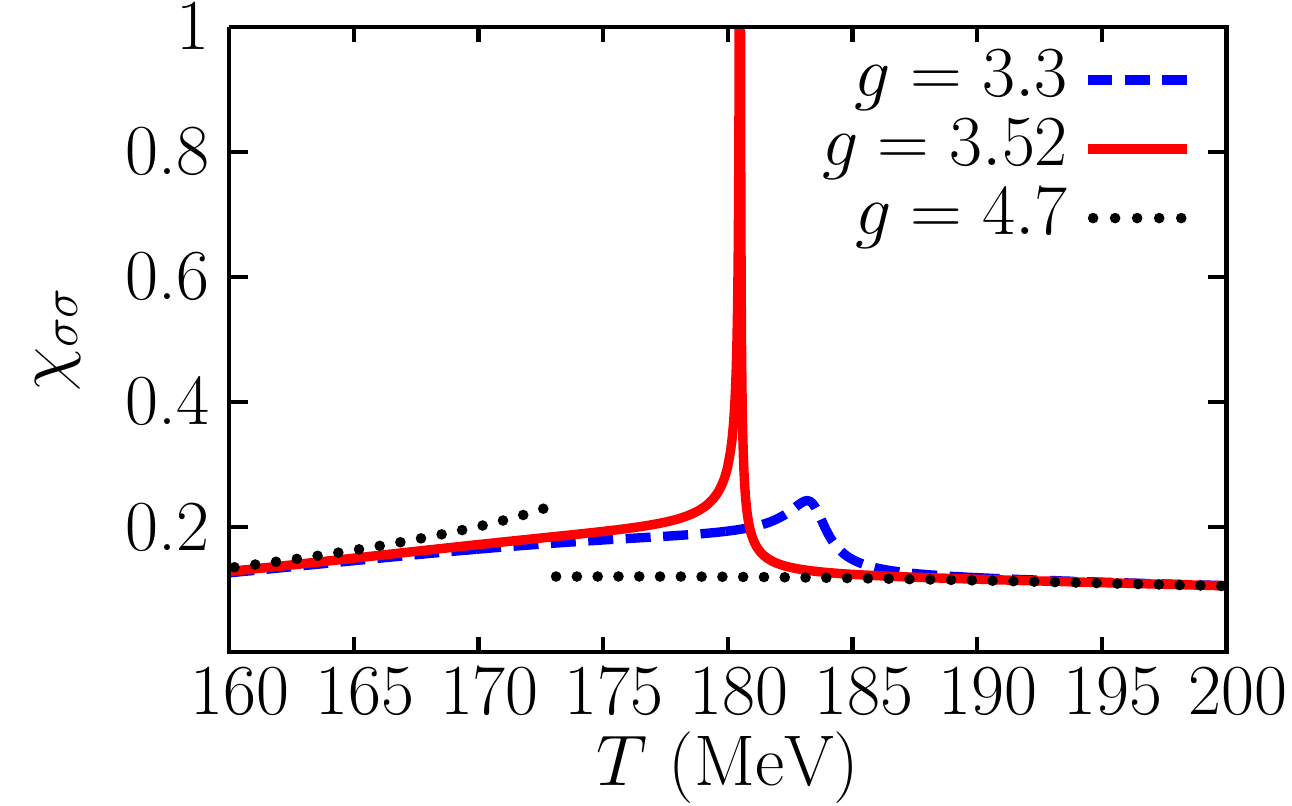}
   }
 \qquad
   \subfloat[\label{fig:susc_polyakov}]{
   \centering
   \includegraphics[scale=0.5]{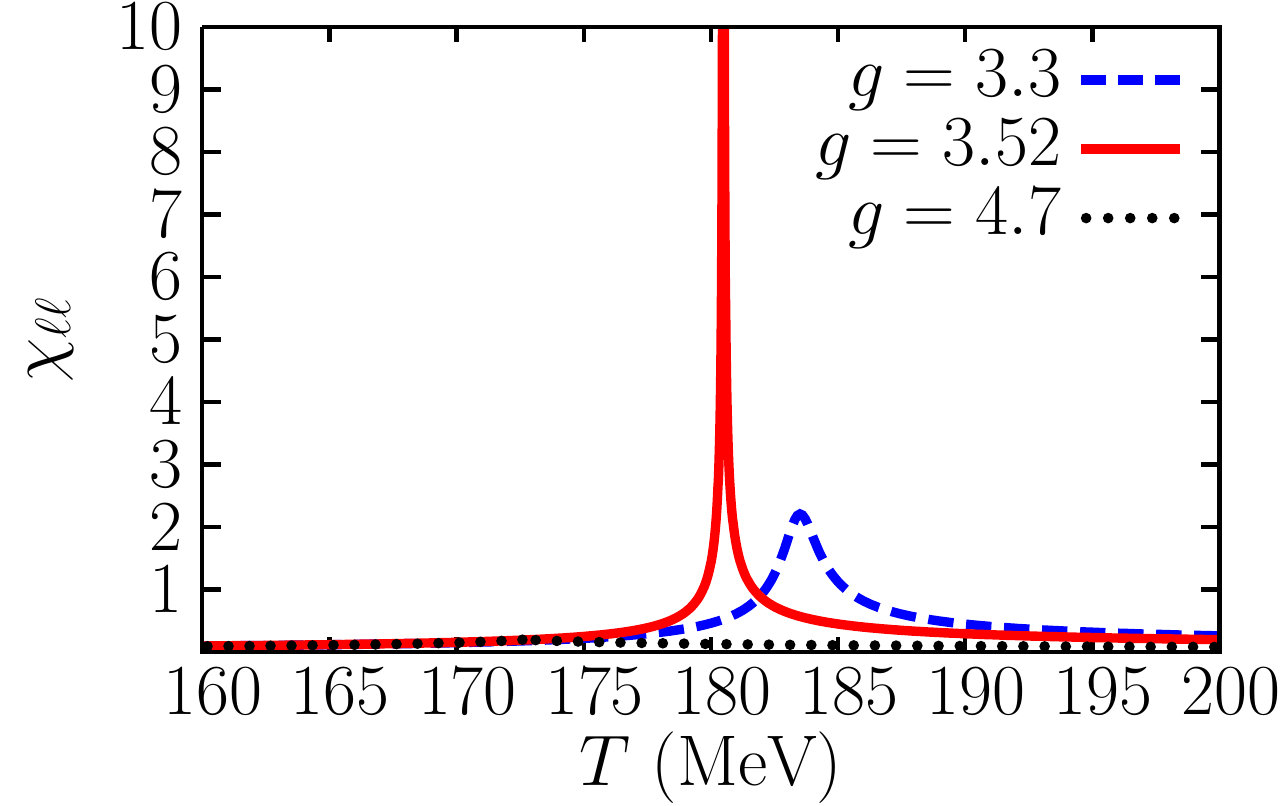}
   }
\caption[.]{\ref{fig:susc_sigma} Chiral susceptibility as a function of temperature for different coupling strengths.
\ref{fig:susc_polyakov} Polyakov loop susceptibility as a function of temperature for different coupling strengths. }
\label{fig:susceptibilities}
\end{figure}

One can quantify this behavior by calculating the chiral and Polyakov loop susceptibilities $\chi_{\sigma\sigma}$ and 
$\chi_{\ell\ell}$. In Fig.~\ref{fig:susceptibilities} they are shown for three different couplings. We find divergent 
susceptibilities for $g=3.52$ indicating a chiral and deconfinement CP. 

Within the two-particle irreducible effective action formalism we self-consistently derived the coupled dynamics for the sigma field and the quark heat bath 
\cite{Nahrgang:2011mg}. We obtained a Langevin equation for the sigma field with temperature dependent 
damping $\eta_{\sigma}$ and stochastic noise term $\xi_{\sigma}$ that are connected via a dissipation fluctuation relation
\begin{eqnarray}
 \partial_{\mu}\partial^{\mu}\sigma+\eta_{\sigma}(T)\partial_t \sigma+\frac{\partial V_{\rm eff}}{\partial\sigma}&=&\xi_{\sigma}~,\\
 \langle\xi_{\sigma}(t,\vec x)\xi_{\sigma}(t',\vec x')\rangle&=&\frac{1}{V}\delta(t-t')\delta(\vec x-\vec x')m_\sigma\eta_{\sigma}\coth\left(\frac{m_\sigma}{2T}\right)~.
\end{eqnarray}
For the Polyakov loop field we deploy a relaxation equation which also contains stochastic noise: 
\begin{eqnarray}
  \eta_{\ell}\partial_t \ell T^2+\frac{\partial V_{\rm eff}}{\partial\ell}&=&\xi_{\ell}~,\\
  \langle\xi_{\ell}(t,\vec x)\xi_{\ell}(t',\vec x')\rangle T^2&=&\frac{1}{V}\delta(t-t')\delta(\vec x-\vec x')2\eta_{\ell} T~.
\end{eqnarray}
Note here that the Polyakov loop is originally defined only in equilibrium and it is not \emph{a priori} clear what the 
correct dynamics are \cite{Ratti:2005jh}. This approach is therefore purely phenomenological. A similar ansatz with an 
additional kinetic term has been pursued in \cite{Dumitru:2001,Dumitru:2002}. The damping coefficient $\eta_{\ell}$ 
is set to a value of $5/{\rm fm}$. Results are sensitive to this choice only in the vicinity of the first-order transition 
temperature \cite{Herold:2013bi}. 

The quarks are propagated via the equations of ideal relativistic fluid dynamics:
\begin{equation}
 \partial_{\mu}T^{\mu\nu}_q=S^{\nu}_{\sigma}+S^{\nu}_{\ell}~,
\end{equation}
with source terms $S^{\nu}_{\sigma}$ and $S^{\nu}_{\ell}$ describing the energy transfer from the fields to the fluid via 
damping. The energy transfer due to stochastic fluctuations needs to be estimated numerically 
\cite{Herold:2013bi,arXiv:1105.1962}.

\section{Numerical results}

\subsection{Equilibration in a box}
\begin{figure}[h]
\centering
  \subfloat[\label{fig:forelax}]{
  \includegraphics[scale=0.5]{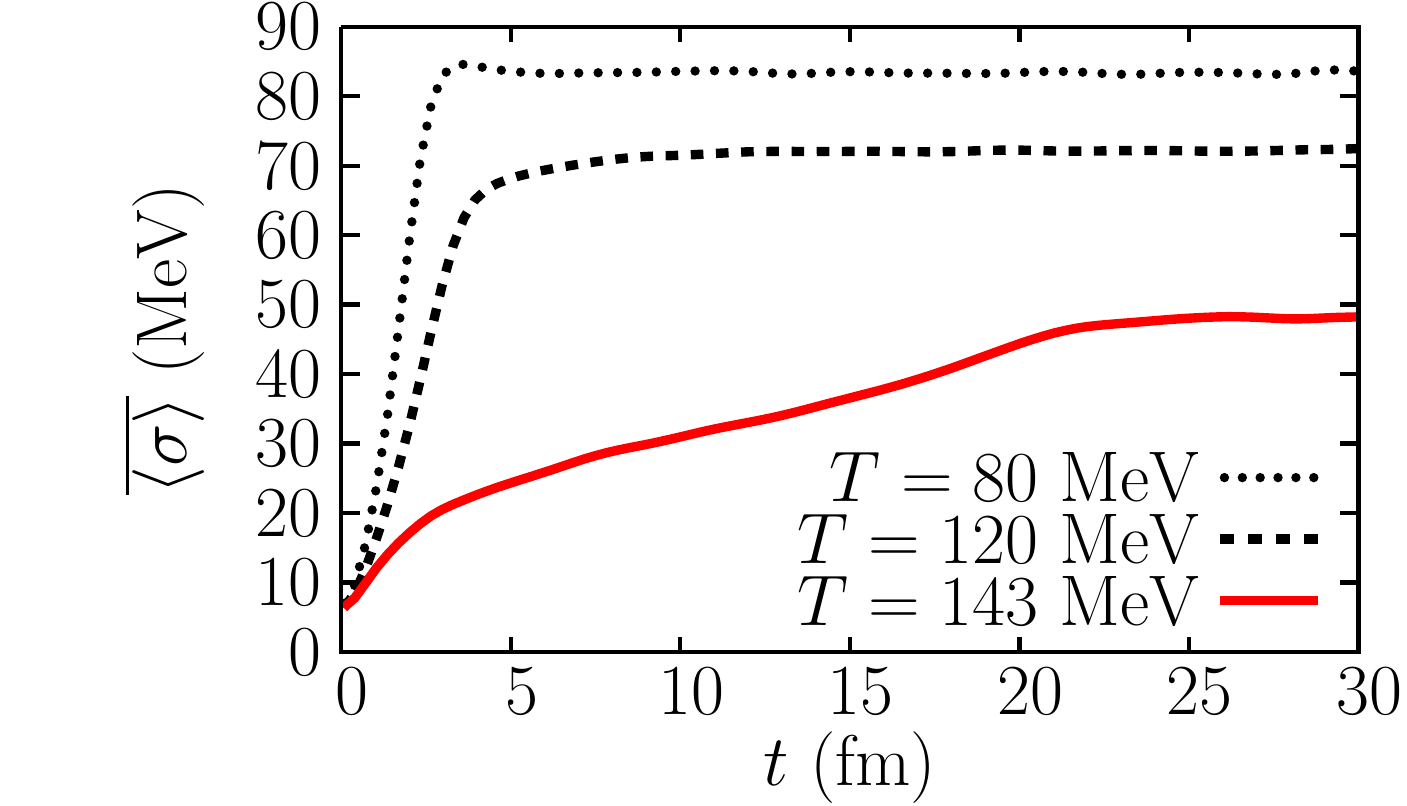}
  }
\quad
  \subfloat[\label{fig:cprelax}]{
  \includegraphics[scale=0.5]{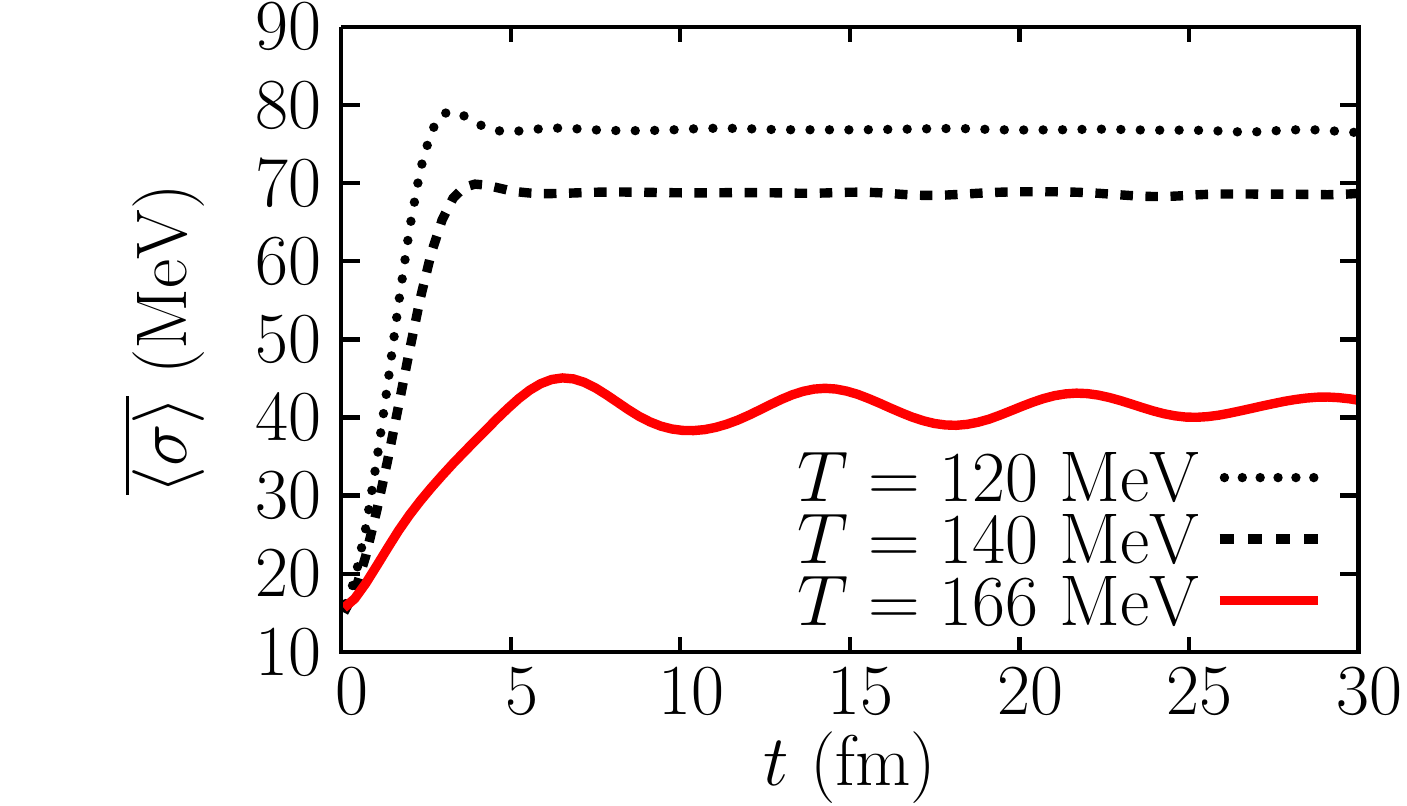}
  }
\caption[.]{\ref{fig:forelax} Equilibration of the sigma field for several quench temperatures $T<T_c$ through the 
first-order transition. The barrier between the minima in the potential increases the relaxation time when the system 
relaxes near $T_c=172.9$~MeV. We chose $T_{\rm ini}=180$~MeV. \ref{fig:cprelax}  Equilibration of the sigma field for several quench 
temperatures $T<T_c$ through the CP. Critical slowing down delays the dynamics and causes oscillations around the flat 
minimum when the system relaxes near $T_c=180.5$~MeV. We chose $T_{\rm ini}=186$~MeV.
Both figures are adopted from \cite{Herold:2013bi}.} 
\label{fig:sigmarelax}
\end{figure}

We study several temperature quenches in a cubic box with periodic boundary conditions. Both fields are initialized at 
some global $T_{ini}>T_c$, with $T_c$ being the respective critical temperature. Then the temperature is quenched to a value 
$T<T_c$ and the energy density and pressure of the quark fluid are calculated. We let the coupled system evolve and relax. 
As pressure gradients are small within this setup, we expect the dynamics to be dominated by the fields. 
The solid red curves in Fig.~\ref{fig:sigmarelax} show the volume and event averaged sigma fiel 
$\overline{\langle\sigma\rangle}$ for equilibration near the transition point for both first-order and CP scenarios. 
At the first-order transition the significant delay in the relaxation time is caused by the large barrier separating the 
degenerate minima. Critical slowing down can be observed near the CP, where the vanishing of $\eta_{\sigma}$ causes 
oscillations and prevents the field from relaxing to its equilibrium state. Similar effects occur in the Polyakov loop 
\cite{Herold:2013bi}. 

\begin{figure}[ht]
\centering
  \subfloat[\label{fig:sig_t24}]{
  \includegraphics[scale=0.45]{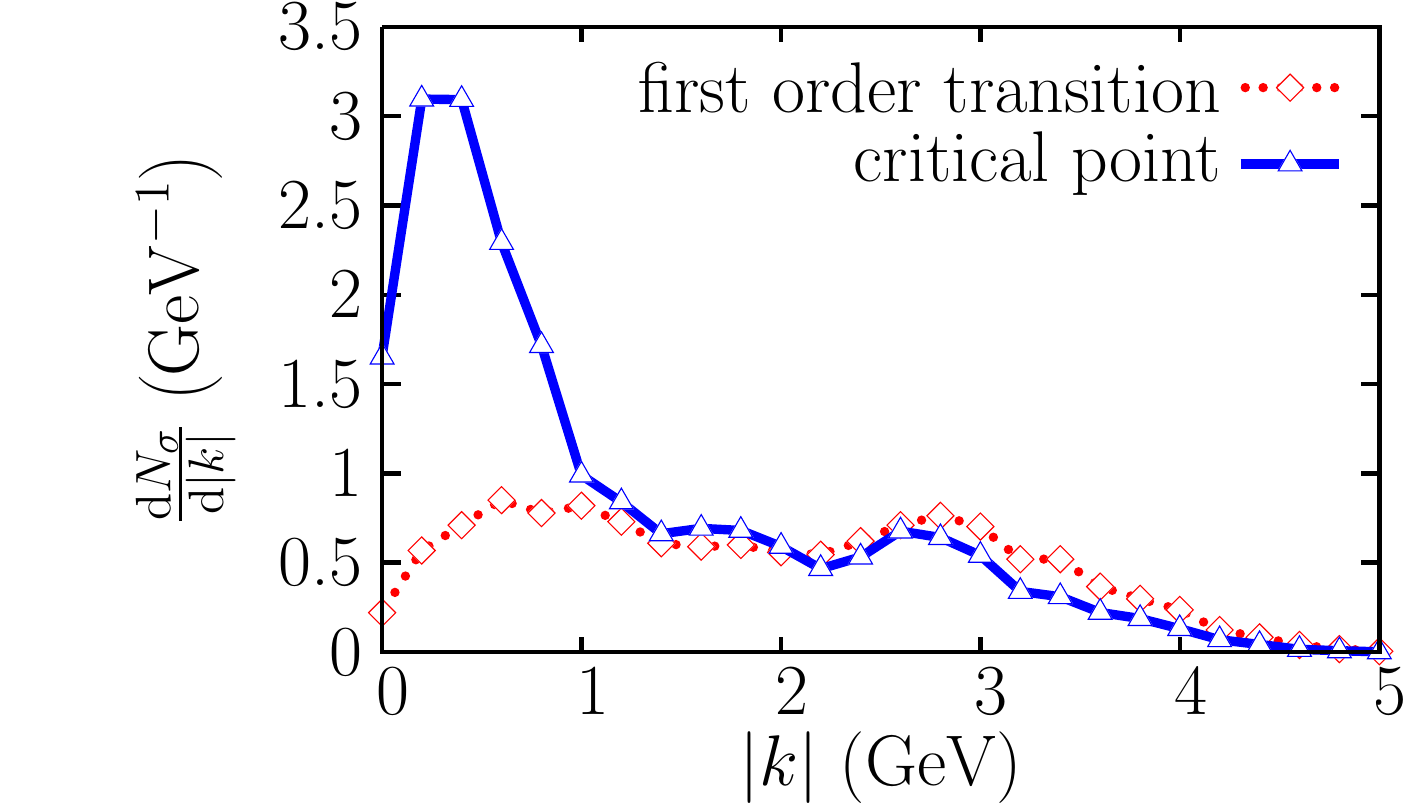}
  }
\qquad
  \subfloat[\label{fig:lo_t24}]{
  \includegraphics[scale=0.45]{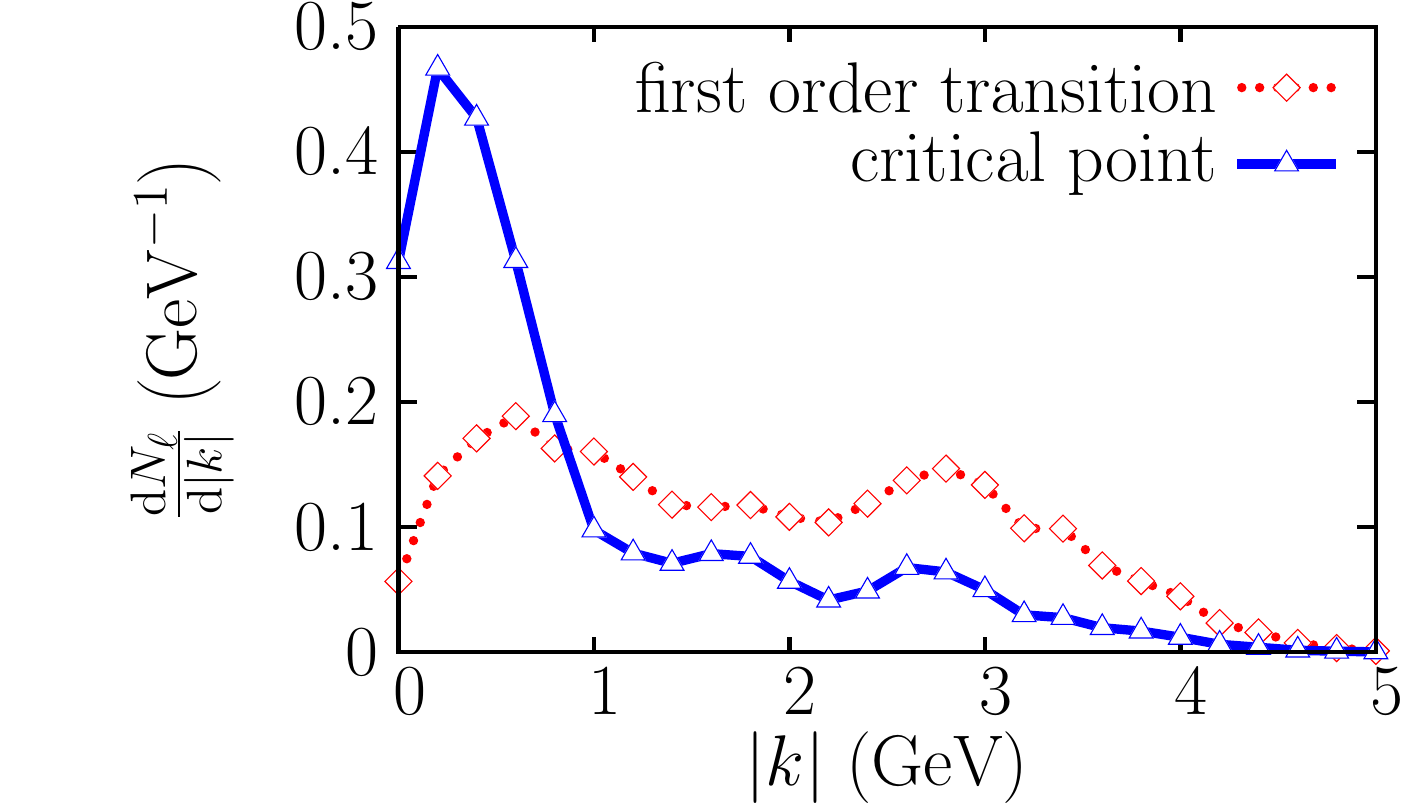}
  }
\caption[.]{\ref{fig:sig_t24} Intensity of sigma fluctuations after equilibration at $t=24$~fm. In the CP scenario we find 
an enhancement of the soft modes. \ref{fig:lo_t24} Intensity of Polyakov loop fluctuations after equilibration at 
$t=24$~fm. In the CP scenario we find an enhancement of the soft modes. Both figures are adopted from \cite{Herold:2013bi}.}
\label{fig:intensity}
\end{figure}

Another critical phenomenon can be observed by studying the intensity of field fluctuations. These are given for the 
sigma and Polyakov loop field as \cite{Herold:2013bi,Abada:1996bw}:
\begin{equation}
\frac{\mathrm d N_{\sigma}}{\mathrm d^3 k}=\frac{\omega_k^2|\delta\sigma_k|^2+|\partial_t\sigma_k|^2}{(2\pi)^3 2\omega_k}~,~~
\frac{\mathrm d N_{\ell}}{\mathrm d^3 k}=T^2 \frac{\omega_k^2|\delta\ell_k|^2+|\partial_t\ell_k|^2}{(2\pi)^3 2\omega_k}~.
\end{equation}
Here $\delta\sigma_k$ and $\partial_t\sigma_k$ are the $k$th Fourier modes of $\delta\sigma=\sigma-\sigma_{\rm eq}$ and 
$\partial_t\sigma$ and $\omega_k$ is the corresponding energy. We compare intensity histograms in the late stage 
of the evolution in the CP and first-order scenario in Fig.~\ref{fig:intensity}. For both order parameter fields we find 
a strong enhancement of long-wavelength modes at the CP compared to an equilibration near the first-order transition point.  

\subsection{Fluid dynamic expansion}
\begin{figure}[ht]
\centering
  \subfloat[\label{fig:sigmadev}]{
  \includegraphics[scale=0.45]{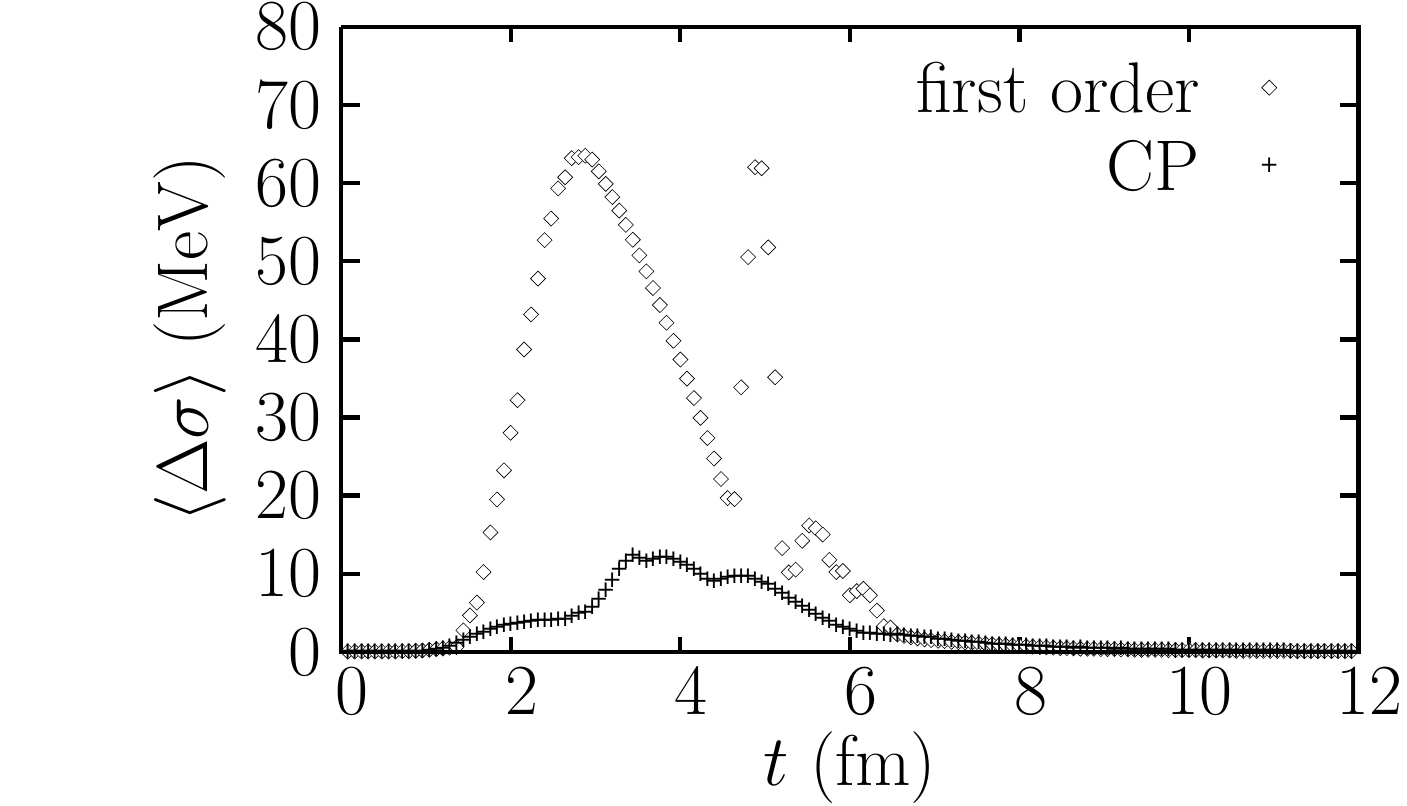}
  }
\qquad
  \subfloat[\label{fig:loopdev}]{
  \includegraphics[scale=0.45]{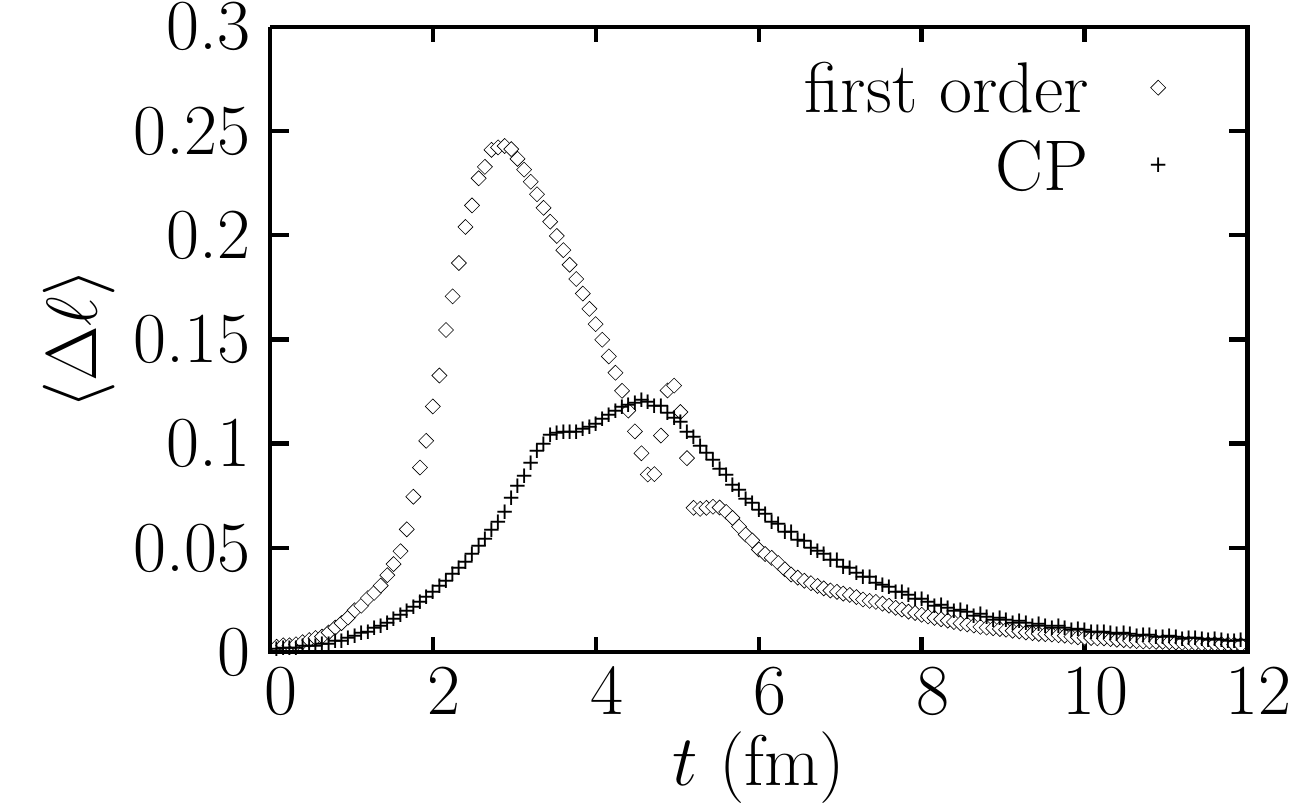}
  }
\caption[.]{Nonequilibrium fluctuations of the sigma field \ref{fig:sigmadev} and Polyakov loop \ref{fig:loopdev} 
are enhanced at the first-order transition compared to the CP scenario.}
\label{fig:noneq}
\end{figure}

To explore the influence of the expansion on the dynamics of the fields, an ellipsoidal region with a temperature 
$T=200$~MeV, above both transition temperatures, is provided as initial state of a fluid dynamic expansion. This is to 
resemble the situation after the collision of two heavy nuclei. Fields and fluid are again set to their respective 
equilibrium values and the system evolves according to full (3+1)-dimensional fluid dynamics. During the expansion 
we observe supercooling and reheating in the first-order transition scenario. This supercooling causes an enhancement 
of nonequilibrium fluctuations 
$\langle\Delta\sigma\rangle=\sqrt{\langle\left(\sigma-\sigma_{\rm eq}\right)^2\rangle}$ and 
$\langle\Delta\ell\rangle=\sqrt{\langle\left(\ell-\ell_{\rm eq}\right)^2\rangle}$ in both order parameters at the 
first-order phase transition, see Fig.~\ref{fig:noneq}. The second bump in the fluctuation strength near $t=6$~fm 
arises when parts of the system cross the transition temperature a second time after reheating.

\subsection{Domain formation at the first-order phase transition}

We now focus on the evolution of a single event to learn more about the transition processes. To achieve this we 
introduce spatial correlations for the stochastic noise fields over volumes $1/m_{\sigma}^3$ and $1/m_{\ell}^3$ to 
obtain a more physical behavior of these fluctuations. 

\begin{figure}[h]
\centering
  \subfloat[\label{fig:sigma_t1}]{
  \includegraphics[scale=0.4, angle=270]{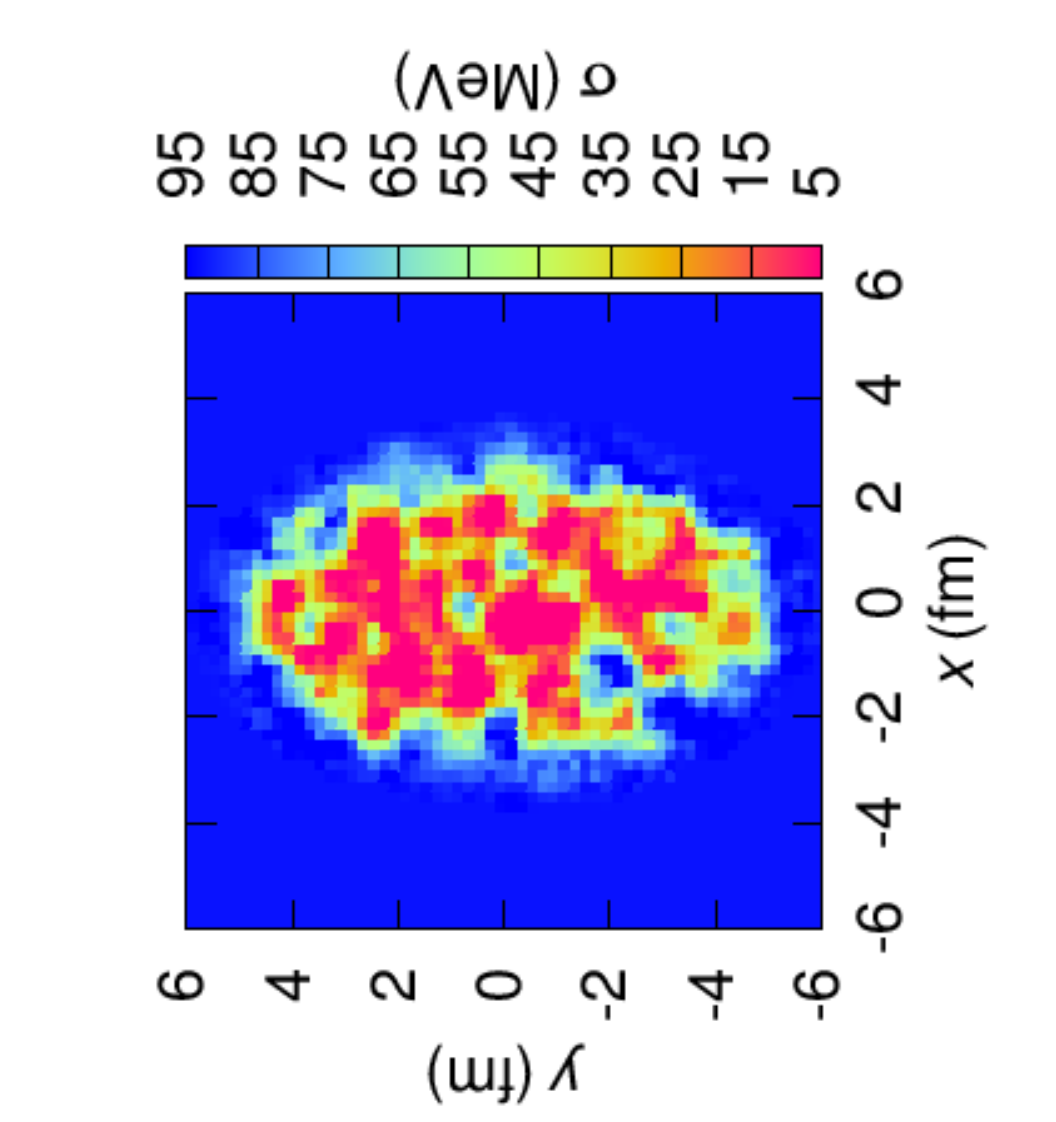}
  }
\quad
  \subfloat[\label{fig:sigma_t4}]{
  \includegraphics[scale=0.4, angle=270]{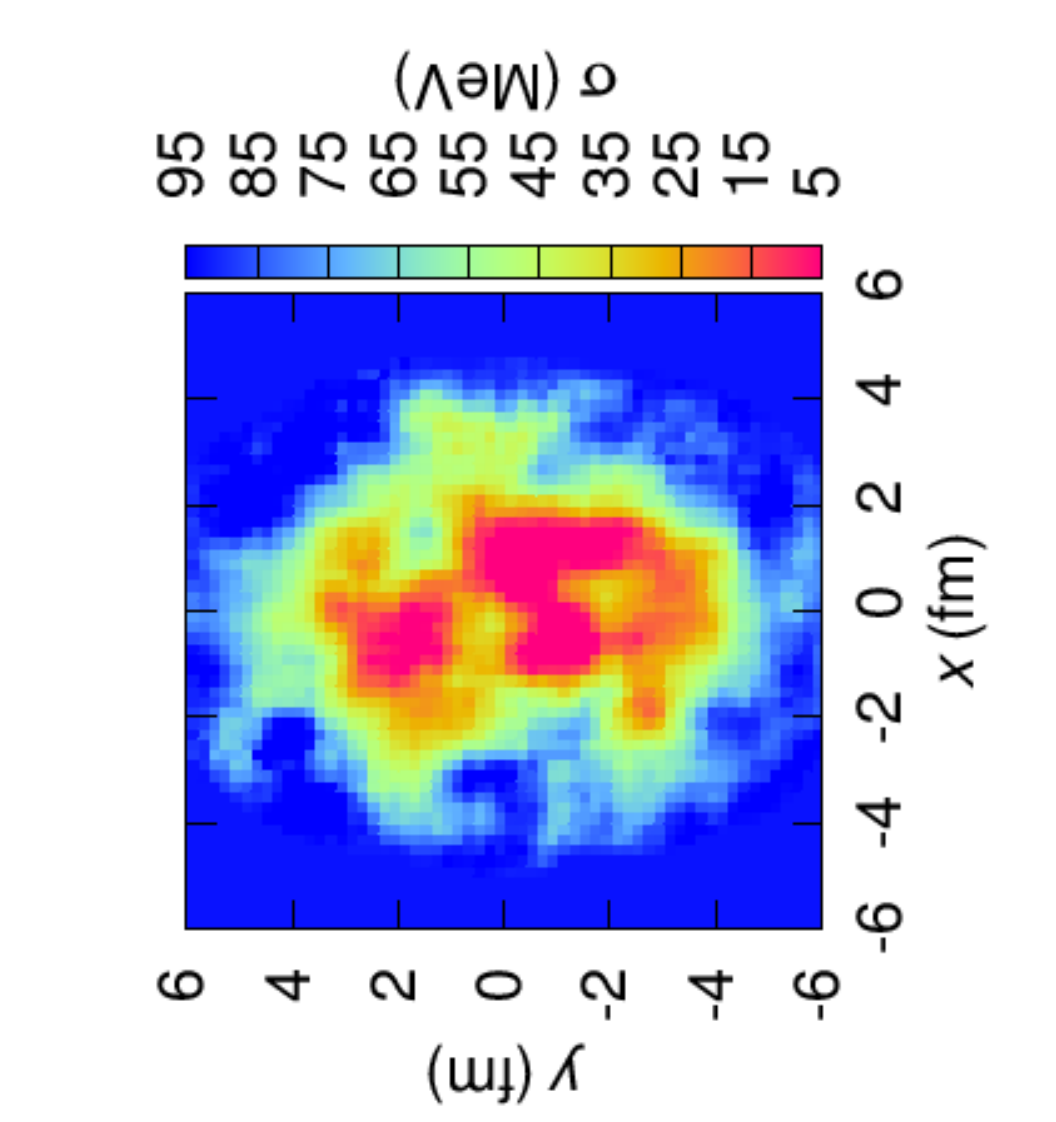}
  }
\quad
  \subfloat[\label{fig:sigma_t7}]{
  \includegraphics[scale=0.4, angle=270]{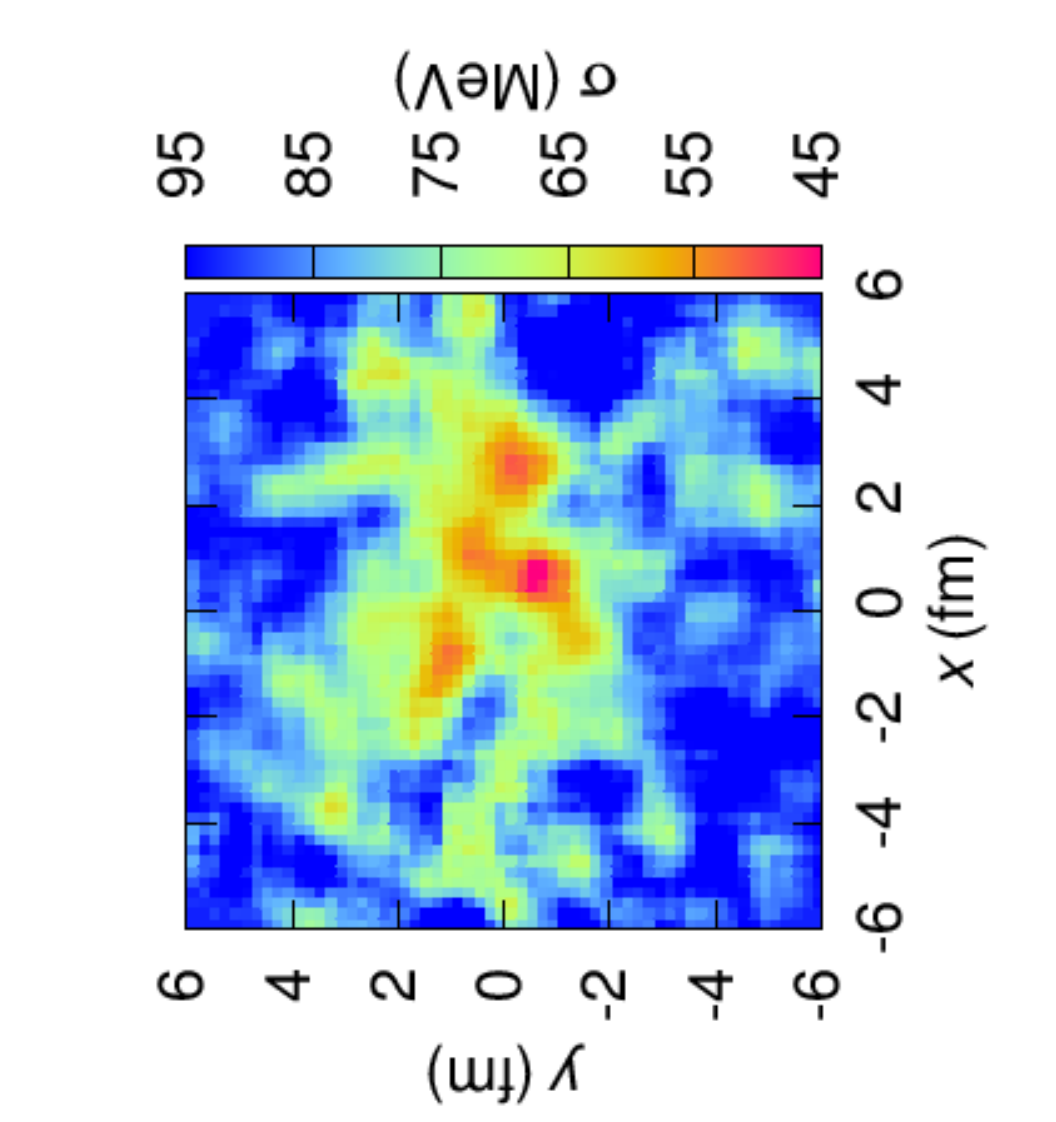}
  }
\caption[.]{Sigma field in the $z=0$ plane for $t=1$~fm \ref{fig:sigma_t1}, $t=4$~fm \ref{fig:sigma_t4}, 
and $t=7$~fm \ref{fig:sigma_t7} during a first-order phase transition. 
Fig.~\ref{fig:sigma_t4} adopted from \cite{Herold:2013bi}.}
\label{fig:mapssigma}
\end{figure}

\begin{figure}[h]
\centering
  \subfloat[\label{fig:loop_t1}]{
  \includegraphics[scale=0.4, angle=270]{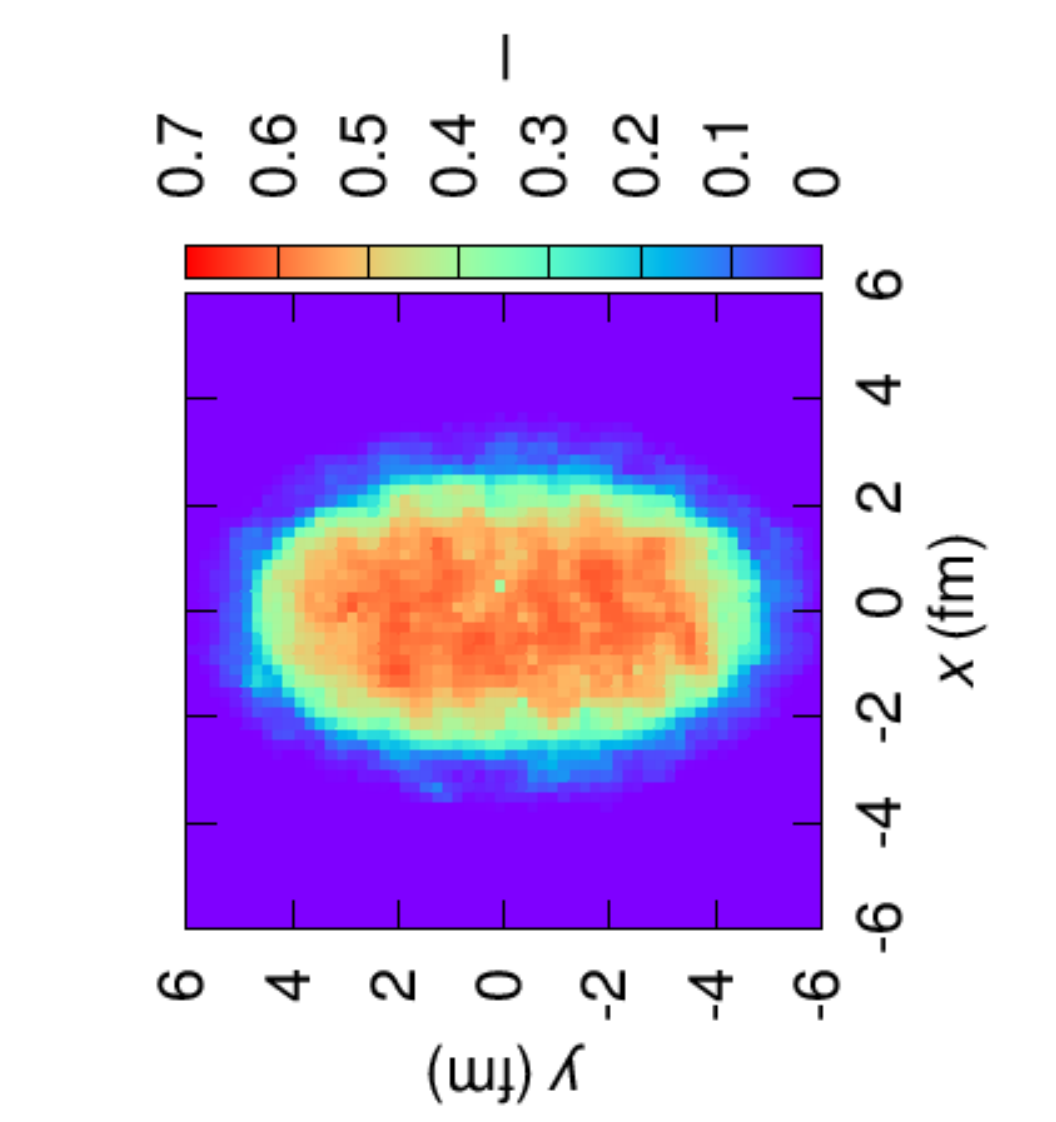}
  }
\quad
  \subfloat[\label{fig:loop_t4}]{
  \includegraphics[scale=0.4, angle=270]{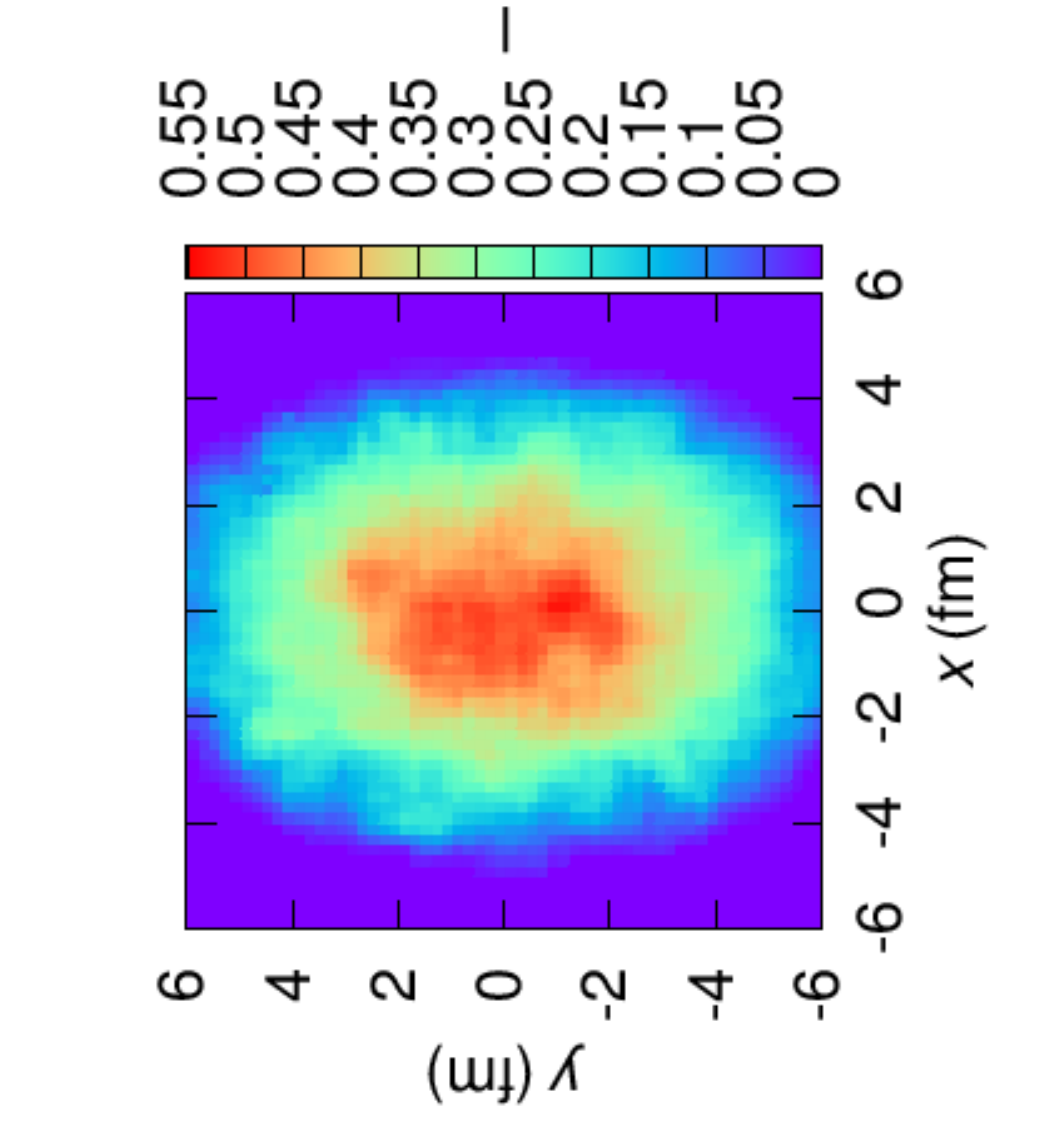}
  }
\quad
  \subfloat[\label{fig:loop_t7}]{
  \includegraphics[scale=0.4, angle=270]{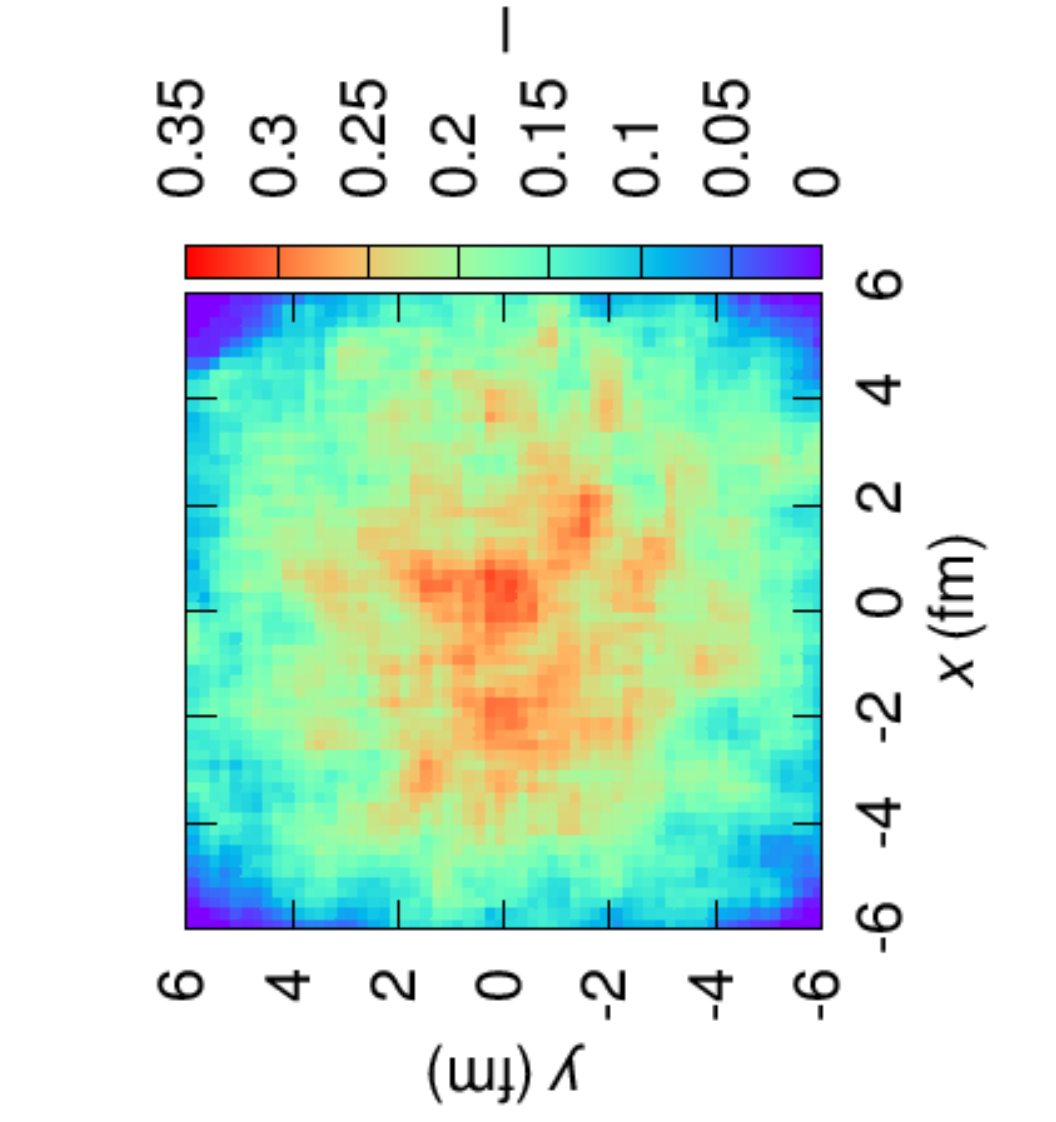}
  }
\caption[.]{Polyakov loop field in the $z=0$ plane for $t=1$~fm \ref{fig:loop_t1}, $t=4$~fm \ref{fig:loop_t4}, 
and $t=7$~fm \ref{fig:loop_t7} during a first-order phase transition. 
Fig.~\ref{fig:loop_t4} adopted from \cite{Herold:2013bi}.}
\label{fig:mapsloop}
\end{figure}

\begin{figure}[h]
\centering
  \subfloat[\label{fig:e_t1}]{
  \includegraphics[scale=0.4, angle=270]{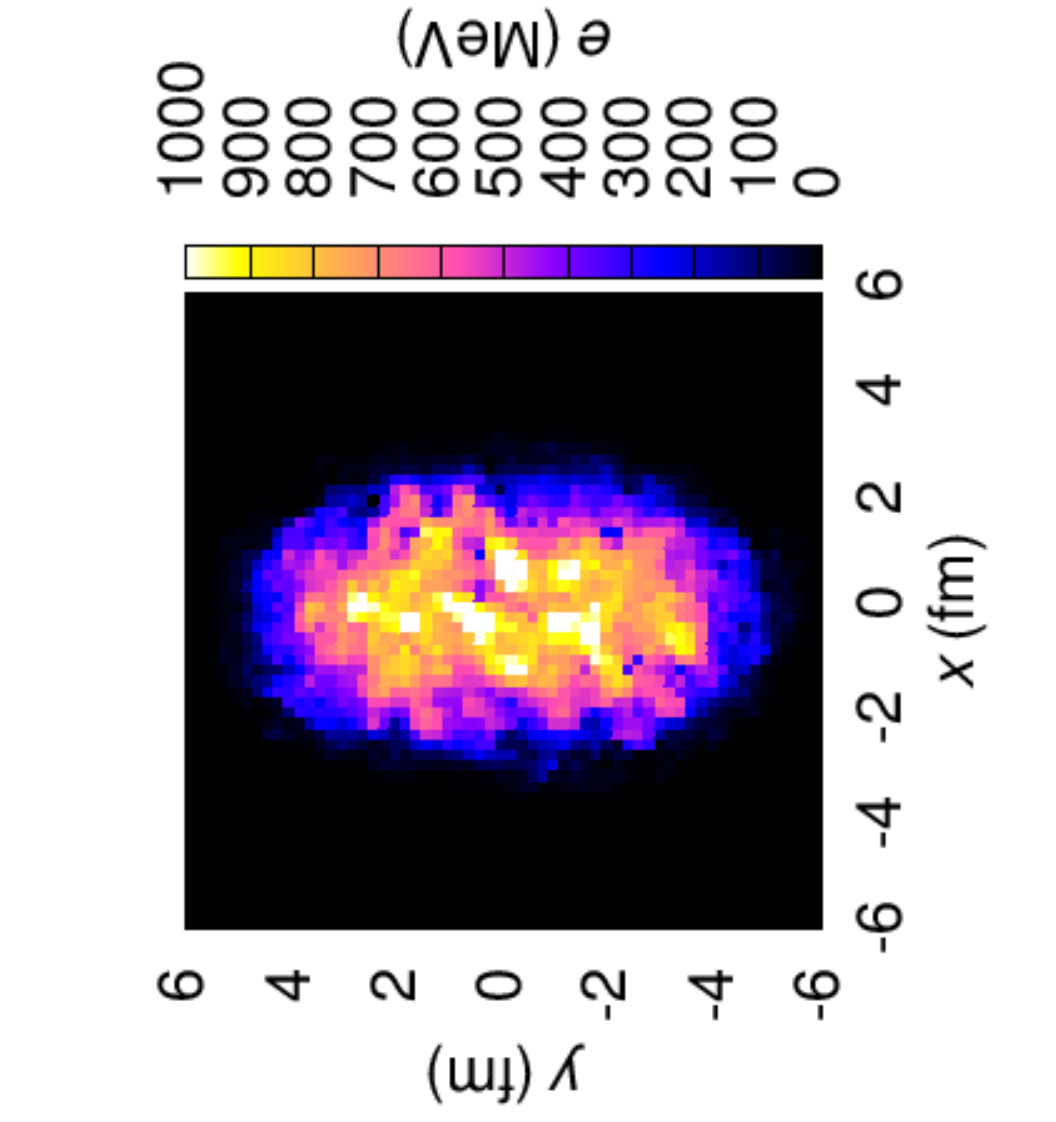}
  }
\quad
  \subfloat[\label{fig:e_t4}]{
  \includegraphics[scale=0.4, angle=270]{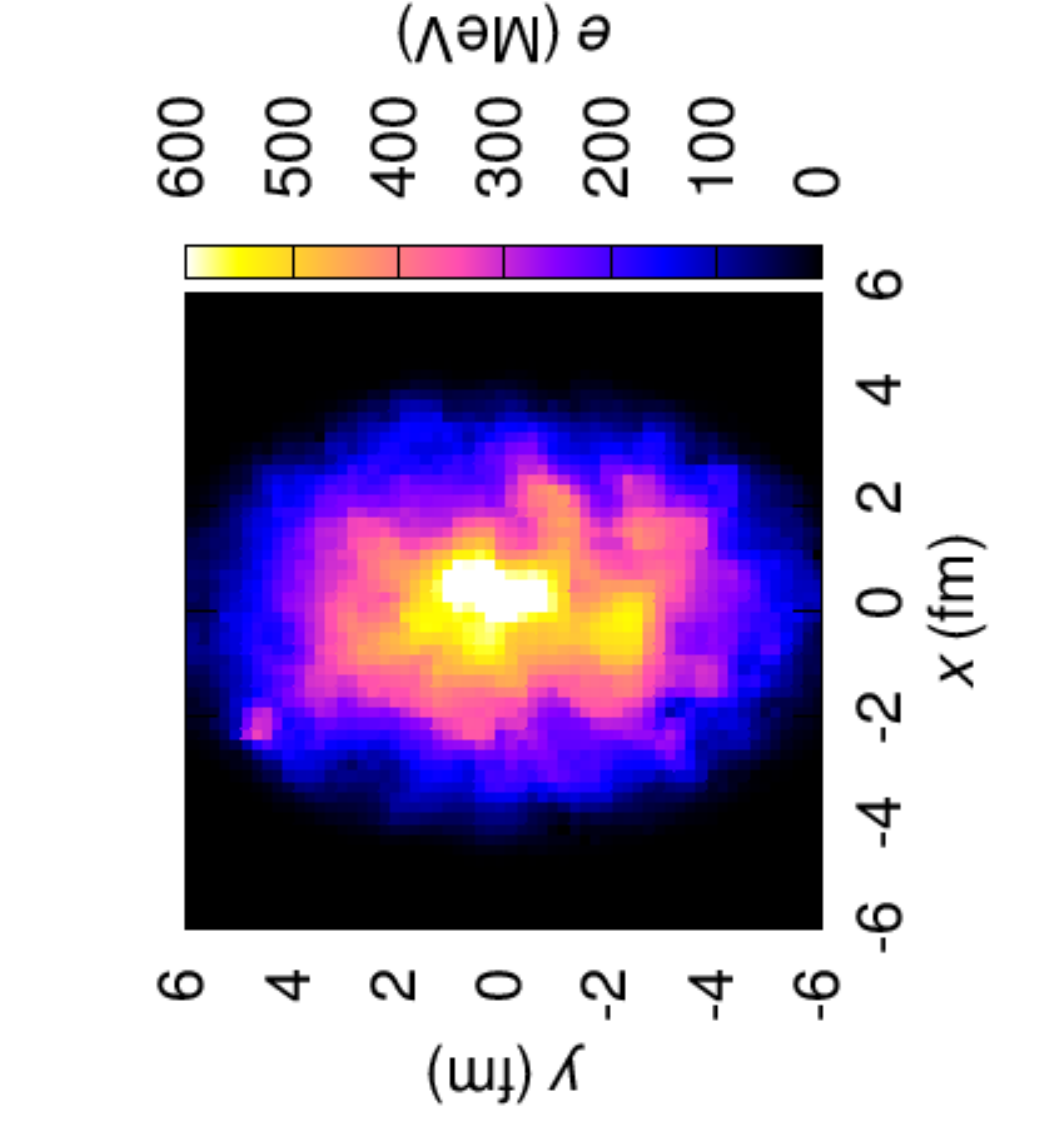}
  }
\quad
  \subfloat[\label{fig:e_t7}]{
  \includegraphics[scale=0.4, angle=270]{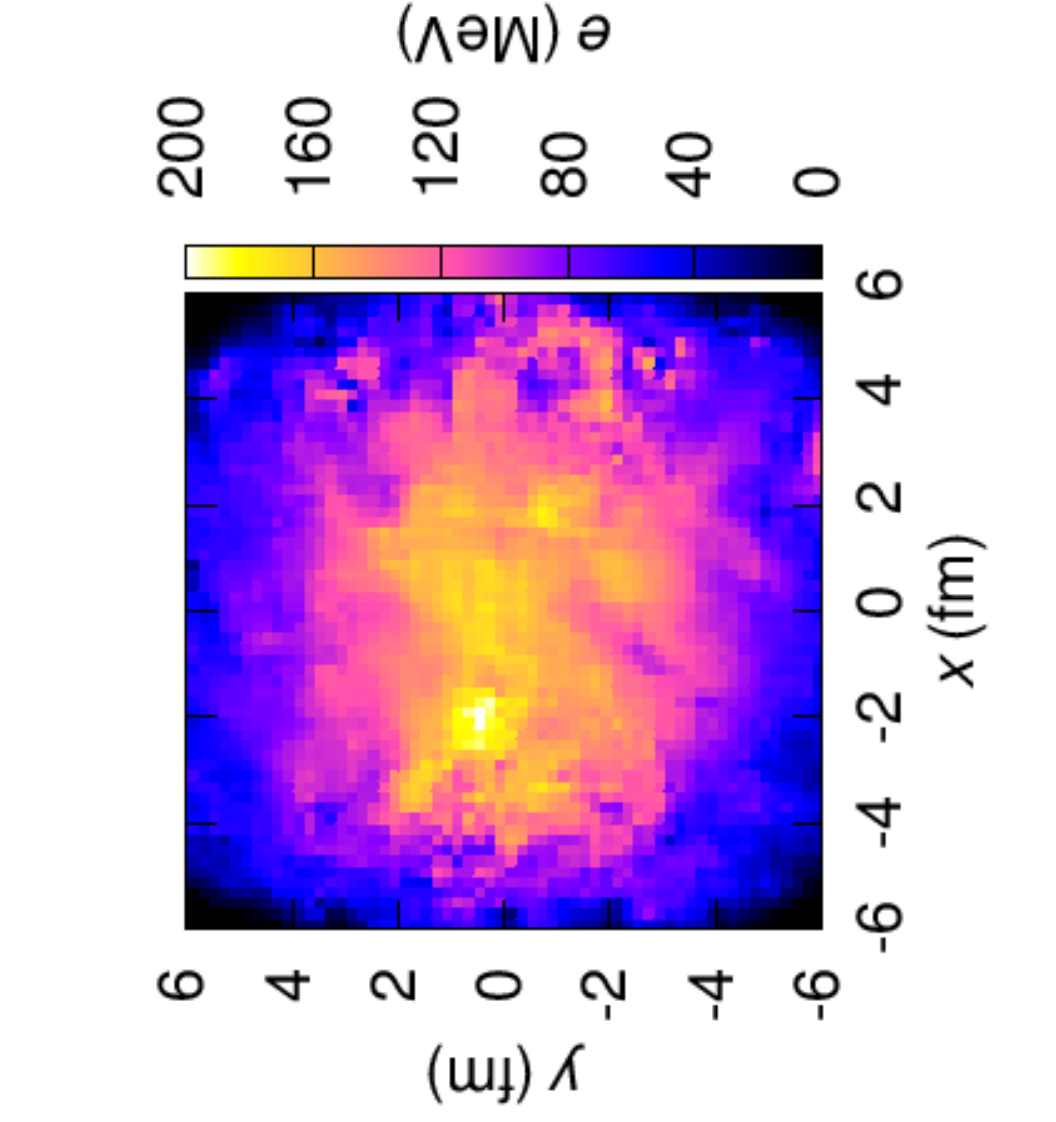}
  }
\caption[.]{Energy density in the $z=0$ plane for $t=1$~fm \ref{fig:e_t1}, $t=4$~fm \ref{fig:e_t4}, 
and $t=7$~fm \ref{fig:e_t7} during a first-order phase transition. 
Fig.~\ref{fig:e_t4} adopted from \cite{Herold:2013bi}.}
\label{fig:mapse}
\end{figure}

We show a slice in the 
transversal $z=0$ plane for the sigma field, Polyakov loop and energy density in Figs. \ref{fig:mapssigma}, 
\ref{fig:mapsloop} and \ref{fig:mapse}, each for early, intermediate and late times in the evolution. We see in  
the order parameters domains of the high- and low-temperature phases coexisting during the transition process. This 
phenomenon is typical for the first-order phase transition and does not occur in evolutions through the CP. It can be best observed in the sigma field, 
but also the Polyakov loop exhibits a bumpy structure during its evolution. This structure then translates to the 
energy density, leading to a significant amount of inhomogeneity and clumping. 

We expect this effect to become even stronger when we go to systems at finite baryon density. This would then provide 
an important experimental signal for the QCD phase transition, e.~g. in non-monotonic multiplicity fluctuations of 
hadrons.

\section{Conclusions}
We presented the extension of nonequilibrium chiral fluid dynamics with a Polyakov loop to include effects of the 
deconfinement phase transition of QCD. We were able to observe typical critical phenomena like critical slowing down and 
the enhancement of soft modes for systems equilibrating near the CP. For an expanding system cooling through the first-order 
phase transition we found evidence for the formation of a supercooled phase leading to subsequent reheating of the fluid. 
As a result, large nonequilibrium fluctuations evolve. For single events, we find significant difference in the 
evolution of fields and fluid between the CP and the first-order scenario. The latter one proceeds through the formation 
of domains in the order parameter fields leading to irregularities in the energy density. As a next step we investigate 
this effect for systems at finite chemical potential to provide relevant signals of the QCD phase transition for 
upcoming experiments at FAIR. 

\acknowledgments
This work was supported by GSI and the Hessian LOEWE initiative Helmholtz International Center for FAIR.


\begin{thebibliography}{99}

\bibitem{Aoki:2006we}
  Y.~Aoki, G.~Endrodi, Z.~Fodor, S.~D.~Katz and K.~K.~Szabo,
  Nature {\bf 443 } (2006) 675-678.

\bibitem{Scavenius:2000qd}
  O.~Scavenius, A.~Mocsy, I.~N.~Mishustin and D.~H.~Rischke,
  Phys.\ Rev.\  C {\bf 64} (2001) 045202.

\bibitem{Stephanov:1999zu}
  M.~A.~Stephanov, K.~Rajagopal and E.~V.~Shuryak,
  Phys.\ Rev.\  D {\bf 60} (1999) 114028.

\bibitem{Stephanov:2008qz}
  M.~A.~Stephanov,
  Phys.\ Rev.\ Lett.\  {\bf 102} (2009) 032301.

\bibitem{Karsch:2010ck}
  F.~Karsch and K.~Redlich,
  Phys.\ Lett.\ B {\bf 695} (2011) 136.

\bibitem{Bleicher:1998ab}
  M.~Bleicher et al., 
  Nuc.\ Phys.\ A {\bf 638} (1998) 391c-394c.

\bibitem{Berdnikov:1999ph}
  B.~Berdnikov and K.~Rajagopal,
  Phys.\ Rev.\  D {\bf 61} (2000) 105017.

\bibitem{Sasaki:2007qh}
  C.~Sasaki, B.~Friman and K.~Redlich,
  Phys.\ Rev.\ D {\bf 77} (2008) 034024.

\bibitem{Steinheimer:2012gc}
  J.~Steinheimer and J.~Randrup,
  Phys.\ Rev.\ Lett.\ {\bf 109} (2012) 212301 .

\bibitem{Herold:2013bi}
  C.~Herold, M.~Nahrgang, I.~Mishustin and M.~Bleicher,
  Phys.\ Rev.\ C {\bf 87} (2013) 014907.

\bibitem{Mishustin:1998eq}
  I.~N.~Mishustin,
  Phys.\ Rev.\ Lett.\  {\bf 82} (1999) 4779.

\bibitem{Steinheimer:2007iy}
  J.~Steinheimer, M.~Bleicher, H.~Petersen, S.~Schramm, H.~Stocker and D.~Zschiesche,
  Phys.\ Rev.\ C {\bf 77} (2008) 034901.

\bibitem{Mishustin:1998yc}
  I.~N.~Mishustin and O.~Scavenius,
  Phys.\ Rev.\ Lett.\  {\bf 83} (1999) 3134.

\bibitem{Paech:2003fe}
  K.~Paech, H.~Stoecker and A.~Dumitru,
  Phys.\ Rev.\  C {\bf 68} (2003) 044907. 

\bibitem{Nahrgang:2011mg}
  M.~Nahrgang, S.~Leupold, C.~Herold and M.~Bleicher,
  Phys.\ Rev.\ C {\bf 84} (2011) 024912.

\bibitem{Nahrgang:2011ll}
  M.~Nahrgang, S.~Leupold and M.~Bleicher,
  Phys.\ Lett.\ B {\bf 711} (2012) 109. 

\bibitem{arXiv:1105.1962}
  M.~Nahrgang, C.~Herold, S.~Leupold, I.~Mishustin and M.~Bleicher,
  arXiv:1105.1962 [nucl-th].

\bibitem{Herold:2013cg}
  C.~Herold, M.~Bleicher and M.~Nahrgang,
  Acta Phys.\ Polon.\ Supp.\  {\bf 5} (2012) 529.

\bibitem{arXiv:0704.3234}
  B.~-J.~Schaefer, J.~M.~Pawlowski and J.~Wambach,
  Phys.\ Rev.\ D\ {\bf 76} (2007) 074023.

\bibitem{Ratti:2005jh}
  C.~Ratti, M.~A.~Thaler and W.~Weise,
  Phys.\ Rev.\ D {\bf 73} (2006) 014019.

\bibitem{Dumitru:2001}
  A.~Dumitru and R.~D.~Pisarski,
  Phys.\ Lett.\ B\ {\bf 504} (2001) 282.

\bibitem{Dumitru:2002}
  A.~Dumitru and R.~D.~Pisarski,
  Nucl.\ Phys.\ A\ {\bf 698} (2002) 444.

\bibitem{Abada:1996bw}
  A.~Abada and M.~C.~Birse,
  Phys.\ Rev.\ D {\bf 55} (1997) 6887.

\end{thebibliography}
\end{document}